\documentclass[]{aa}
\usepackage{txfonts}
\usepackage{xspace,colortbl,graphicx,natbib}

\begin{document}

\title{Doppler shift oscillations in solar spicules}
\author{T.V. Zaqarashvili, E. Khutsishvili, V. Kukhianidze \and G. Ramishvili}

\institute{Georgian National Astrophysical Observatory (Abastumani Astrophysical Observatory), Kazbegi Ave. 2a, Tbilisi 0160, Georgia, \\
\email{temury@genao.org}}

\offprints{T.V. Zaqarashvili}

\date{Received / Accepted }

\abstract{Consecutive height series of H${\alpha}$ spectra in solar
limb spicules taken on the 53 cm coronagraph of Abastumani
Astrophysical Observatory at the heights of 3800-8700 km above the
photosphere have been analyzed.}{The aim is to observe oscillatory
phenomena in spicules and consequently to trace wave propagations
through the chromosphere.}{The Discrete Fourier Transform analysis
of H${\alpha}$ Doppler shift time series constructed from the
observed spectra at each height is used.}{Doppler velocities of
solar limb spicules show oscillations with periods of 20-55 and
75-110 s. There is also the clear evidence of 3-min oscillations at
the observed heights.}{The oscillations can be caused by wave
propagations in thin magnetic flux tubes anchored in the
photosphere. We suggest the granulation as a possible source for the
wave excitation. Observed waves can be used as a tool for {\it
spicule seismology}; the magnetic field strength in spicules at the
height of $\sim$ 6000 km above the photosphere is estimated as
$12-15$ G.}


\titlerunning{Oscillations in solar spicules}
\authorrunning{Zaqarashvili et al.}

\maketitle

\section{Introduction}

It is widely believed that the energy source responsible to heat the
coronal plasma up to 1 MK is located in denser and dynamic
photosphere. Chromospheric structured magnetic fields may "guide" an
energy towards the corona, which can be dissipated there leading to
heat ambient plasma. One of most plausible mechanisms of the energy
transport is due to magnetohydrodynamic (MHD) waves. The waves can
be generated in photospheric magnetic tubes by buffeting of granular
motions (Roberts 1979; Spruit 1981). Then they may propagate along
the chromospheric magnetic field, penetrate into the corona and
deposit the energy into heat. Therefore observations of oscillatory
motions in the chromosphere is a key test for the wave heating
theory.

Most pronounced features of the chromosphere in quiet Sun regions
are spicules;  jet-like limb structures observed mainly in
H${\alpha}$ line (Beckers 1972). They are concentrated between
supergranule cells and thus probably are formed in regions of
intense magnetic field concentrations, although the formation
mechanism is not yet known (Sterling 2000; but see Roberts 1979 and
De Pontieu et al. 2004). On the-other-hand, spicules may rise along
the magnetic tubes, which at the same time guide MHD waves from the
photosphere into the corona. Therefore the wave propagation in the
chromosphere may be traced through oscillatory dynamics of spicule
plasma.

Oscillations in spicules have been observed mostly with $\sim$5 min
period (Kulidzanishvili \& Zhugzhda 1983; De Pontieu et al. 2003;
Xia et al. 2005), which probably are connected with global p-modes.
On the-other-hand, oscillations in spicules with shorter period
($\sim$ 1 min) have been reported by Nikolsky \& Platova (1971) as
periodic transversal displacements of spicule axes at one particular
height. Recent observations of higher frequency waves in the "green"
coronal line during the August 1999 total solar eclipse (Katsiyannis
et al. 2003), in the Fe I 5434 $\AA$ line by German Vacuum Tower
Telescope on Tenerife (Wunnenberg et al. 2002) and in the transition
region spectral lines by TRACE (Deforest 2004; de Wijn et al. 2005)
indicate to the significant power at the high frequency branch of
oscillations in almost whole solar atmosphere. This further
stimulates to search of short period oscillations in spicules. These
short period waves may have a significant input in the chromospheric
and coronal heating. Note, that the excitation of short period waves
(10-20 s) in photospheric magnetic flux tubes has been proposed
recently by Zaqarashvili and Roberts 2002).

Recently Kukhianidze et al. (2006, hereinafter paper I) reported
periodic spatial distributions of Doppler velocities with height
through spectroscopic analysis of H${\alpha}$ height series in solar
limb spicules (at the heights of 3800-8700 km above the
photosphere). They found that nearly 20$\%$ of measured height
series show a periodic spatial behaviour with $\sim$ 3500 km. This
spatial periodicity in Doppler velocities was explained as a
signature of kink wave propagation in spicules. Wave periods were
estimated as 35-70 s based on the expected kink speed in the
chromosphere (50-100 km s$^{-1}$). The observed wave length was
suggested to be shorter ($\sim$ 800-1000 km) at the photospheric
level due to the decrease of kink speed. Estimated wave length at
the photospheric level is comparable to spatial dimensions of
granular cells, therefore the granulation was suggested as a
possible source for the wave excitation. Observations of the waves
may be related to the solution of coronal heating problem.
Therefore, for further searching of oscillatory motions in the
chromosphere, we performed complete analysis of H${\alpha}$ series,
preliminary results of which were presented in the paper I. Here we
study a temporal dynamics of consecutive H${\alpha}$ spectra with a
time interval of $\sim$ 7-8 s between consecutive measurements at
fixed heights, which cover almost whole life time of spicules (7-15
min).

\begin{figure}
\includegraphics[width=8.5cm]{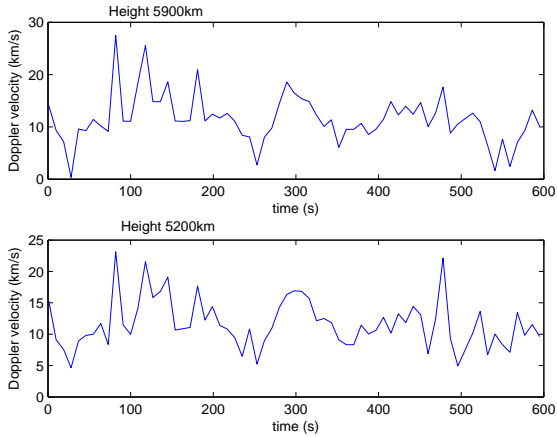}
\caption{Doppler velocity time series at third ($\sim$ 5200 km) and
fourth ($\sim$ 5900 km) heights in the spicule I. The time interval
between consecutive measurements is $\sim$ 8 s. \label{fig:epsart}}
\end{figure}

We show a spectroscopic evidence of wave propagation in the
chromosphere in terms of Doppler velocity oscillations in time at
particular heights of solar limb spicules.

\section{Observation and data analysis}

Observations have been carried out on the big (53 cm) coronagraph of
Abastumani Astrophysical Observatory (instrumental spectral
resolution and dispersion in H${\alpha}$ are 0.04 {\AA} and 1
{\AA}/mm correspondingly) in September 26, 1981 at the solar limb as
height series beginning at 3800 km height from the photosphere and
upwards (for details, see Khutsishvili 1986). Chromospheric
H${\alpha}$ line was used to observe solar limb spicules at 8
different heights. The distance between neighboring heights was
1$^{\prime \prime}$ (which was the spatial resolution of
observations), thus the distance $\sim$ 3800-8700 km above the
photosphere was covered. The exposure time was 0.4 s at four lower
heights and 0.8 s at higher ones. The total time duration of each
height series was 7 s. The consecutive height series begins
immediately. Therefore continuous time series of H${\alpha}$ spectra
with interval of $\sim$ 7-8 s between consecutive measurements at
each height can be constructed. The time series cover almost whole
life time (from 7 to 15 min) of several spicules.

\begin{figure}
\includegraphics[width=8.5cm]{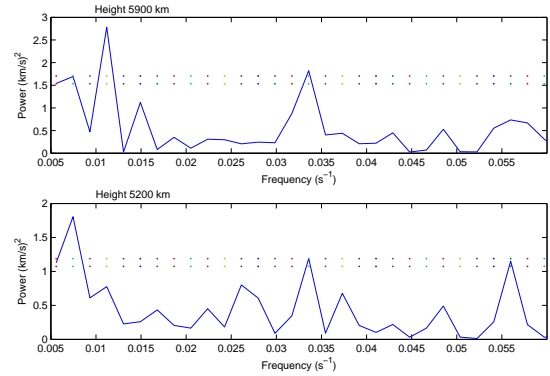}
\caption{Power spectra of Doppler velocity oscillations in the
spicule I at the heights of 5200 km (lower panel) and 5900 (upper
panel) correspondingly. The dotted lines in both plots show 95.5$\%$
and 98$\%$ confidence levels. Here is the clear evidence of
oscillations with the periods of 180 and 30 s at both heights.
\label{fig:epsart}}
\end{figure}

Each H${\alpha}$ line profile from the time series was fitted to a
Gaussian. Then temporal variations of the Gaussian center with
respect to the photospheric reference line (4371 {\AA}) have been
studied at each heights for four different spicules. We have
calculated the Doppler shifts, and consequently Doppler velocities,
with 7-8 s interval at each heights in all spicules. Fig.1 shows the
Doppler velocity time series at third ($\sim$ 5200 km) and fourth
($\sim$ 5900 km) heights in one of spicules. Then the spectral
analysis of the time series at all heights has been carried out with
the Discrete Fourier Transform (DFT). DFT enables to reveal
oscillation periods with 20 - 250 s; the shorter periods are
restricted due to an interval between consecutive measurements and
the longer periods are restricted due to life times of spicules.

\begin{figure}
\includegraphics[width=8.5cm]{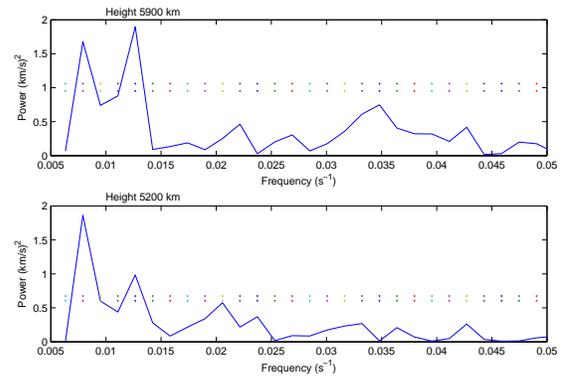}
\caption{Power spectra of Doppler velocity oscillations in the
spicule II at the heights of 5200 km (lower panel) and 5900 km
(upper panel). The dotted lines show 95.5$\%$ and 98$\%$ confidence
levels. This spicule shows the oscillations with the periods of 120
and 80 s at both heights. \label{fig:epsart}}
\end{figure}

\section{Results}

In this section, we present the results of DFT in four different
spicules separately. Best fit of H${\alpha}$ line profiles to a
Gaussian was found at third ($\sim$ 5200 km) and fourth ($\sim$ 5900
km) heights. The fit was relatively poor at lower and higher
heights. Therefore calculated Doppler shifts are more confident at
third and fourth heights. For this reason, we first show the results
of DFT for these heights, then turn to general oscillatory phenomena
at all heights.

\subsection{Spicule I}

Fig.2 shows the power spectra of Doppler velocity oscillations in
the spicule I at the heights of 5200 km (below) and 5900 km (up).
The dotted lines in both plots show 95.5$\%$ and 98$\%$ confidence
levels respectively.

Most pronounced periods at the height of 5200 km are 180, 30 and 17
s (Fig.2, lower panel). However, 17 s period is probably suspicious
as it is closer to the time interval between consecutive
measurements ($\sim$ 7-8 s). So the oscillations of Doppler velocity
with the periods of 180 and 30 s are more confident at this height.
The upper panel shows the power spectrum at the height of 5900 km.
Here the most pronounced periods are 180, 90 and 30 s. So the
spicule oscillates with the periods of 30 and 180 s at both heights.
The oscillation with the period of 90 s is also seen but preferably
at the higher height (but note the small peak at the lower height as
well).

\begin{figure}
\includegraphics[width=8.5cm]{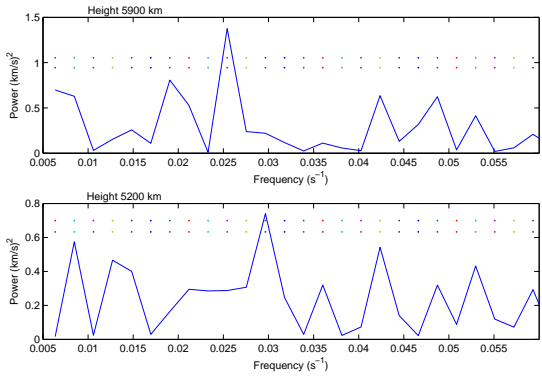}
\caption{Power spectra of Doppler velocity oscillations in the
spicule III at the heights of 5200 km (lower panel) and 5900 km
(upper panel). The dotted lines show 95.5$\%$ and 98$\%$ confidence
levels. This spicule oscillates with $\sim$ 35 s period at both
heights. \label{fig:epsart}}
\end{figure}

\subsection{Spicule II}

Fig.3 shows the power spectra of Doppler velocity oscillations in
the spicule II at the heights of 5200 km (lower panel) and 5900 km
(upper panel). The dotted lines again show 95.5$\%$ and 98$\%$
confidence levels.

We see two clear oscillation periods of 120 and 80 s at both
heights. Both periods are above 98$\%$ confidence level. There is
some evidence of $\sim$ 50 s period at the height of 5200 km, but
just below of 95.5$\%$ confidence level.

\subsection{Spicule III}

Fig.4 shows the power spectra of Doppler velocity oscillations in
the spicule III at the heights of 5200 km (lower panel) and 5900 km
(upper panel).

The spicule shows the oscillations with 37 s period at the height of
5200 km and with 35 s period at the height of 5900 km. Hence, the
spicule oscillates with the period of $\sim 35$ s at both heights.

\subsection{Spicule IV}

Fig.5 shows the power spectra of Doppler velocity oscillations in
the spicule IV at the heights of 5200 km (lower panel) and 5900 km
(upper panel).

Clear oscillation periods in this spicule at both heights are $\sim$
110 and $\sim$ 40 s. There is the evidence of oscillations with 30
s, but just below of 95.5$\%$ confidence level.

\begin{figure}
\includegraphics[width=8.5cm]{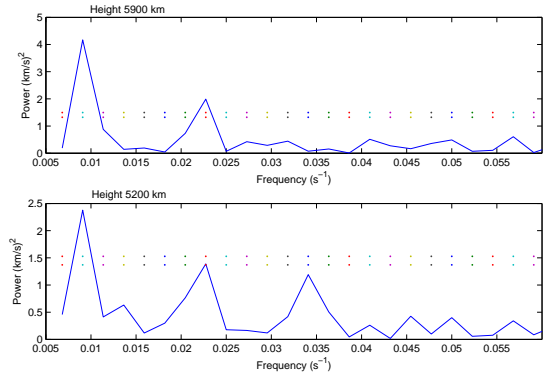}
\caption{Power spectra of Doppler velocity oscillations in the
spicule IV at the heights of 5200 km (lower panel) and 5900 km
(upper panel). The dotted lines show 95.5$\%$ and 98$\%$ confidence
levels. The periods of $\sim$ 110 and $\sim$ 40 s are most
pronounced at both heights. \label{fig:epsart}}
\end{figure}

\subsection{All oscillatory periods above 95.5$\%$ confidence level}

Now let present the results of DFT for all 32 time series; i.e. at 8
different heights in 4 different spicules.

\begin{figure}
\includegraphics[width=8.5cm]{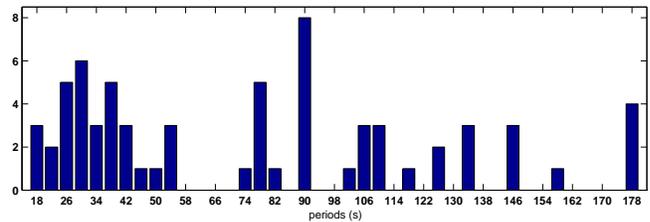}
\caption{Histogram of all oscillation periods which are above
95.5$\%$ confidence level. The horizontal axis shows the oscillation
periods in seconds, while the vertical axis shows the number of
corresponding periods.  \label{fig:epsart}}
\end{figure}

Fig.6 shows a histogram of all oscillation periods which are above
95.5$\%$ confidence level. The histogram reveals interesting
properties of oscillatory periods in spicules. Almost half of the
oscillatory periods are located in the range of 18-55 s, which have
been suggested by Kukhianidze et al. in the paper I. Another
interesting range of oscillatory periods is 75-110 s, with the clear
peak at the period of 90 s. Note, the interesting peak at 178 s
period as well, which is the clear evidence of well known 3 min
oscillations.

In order to show spatial locations of oscillations we plot a Fourier
power (expressed in confidence levels) as a function of frequency
and heights for spicules II (Fig. 7, upper panel) and IV (lower
panel). There is the clear evidence of persisted oscillations along
whole length of both spicules. The plot of the spicule II shows the
long white feature (feature A) located just above the frequency 0.01
s$^{-1}$. This is the oscillation with the period of $\sim$ 80 s
found at third and fours heights (see subsection 3.2). The
oscillation persists along whole spicule and is a signature of
either standing or propagating wave pattern. The most pronounced
feature (feature B) on the plot of the spicule IV is long light
trend located just above the frequency of 0.02 s$^{-1}$ and
persisted along almost whole spicule. This is the oscillation with
period of 44 s (the same period was found at third and fourth
heights; see subsection 3.4). Thus there is the wave pattern with
$\sim$ 40-45 s period in the spicule IV. However, it must be
mentioned again that only the oscillations at third and fourth
heights have high confidence due to good fit of line profile to a
Gaussian.

\section{Discussion}

Time series of H${\alpha}$ spectra in solar limb spicules show the
clear evidence of Doppler shift oscillations, which probably are
caused due to oscillations in line of sight velocity. Spicules have
almost vertical direction at the solar limb, therefore the velocity
is probably transversal to spicule axes. However, longitudinal
oscillations also can not be ruled out if spicule axes are tilted
from the vertical. It is clear that the oscillations in Doppler
velocity can be caused due to wave propagations in spicules.

\subsection{Wave propagation in spicules}

Photospheric granulation is often suggested as a source for wave
excitations in anchored thin magnetic tubes (Roberts 1979; Spruit
1981; Hollweg 1981; Hasan \& Kalkofen 1999; Mishonov et al. 2007).
The waves may propagate along the tubes towards the corona carrying
energy and momentum. The tubes may guide spicule material at the
same time. Therefore the wave propagation in the chromosphere may be
traced through spicule dynamics. Magnetic tubes may guide three
different types of waves: kink, sausage and torsional Alfv\'en
waves. Some of these waves may cause the observed Doppler shift
oscillations. Torsional Alfv\'en waves in thin tubes may lead to
periodic non-thermal broadenings of spectral lines, but not to
Doppler shift oscillations (Zaqrashvili 2003; Zaqarashvili and
Murawski 2007). Sausage waves cause oscillations mainly in a line
intensity due to density variations. However, the longitudinal
velocity field of sausage waves may lead to the Doppler shift
variations if tube axes are significantly tilted from the vertical.
But the main contributor into Doppler shift oscillations at the
solar limb probably are kink waves, which oscillate transverse to
the tube axis. We argue that the back and forth transversal motions
of vertical tube axis at the solar limb due to the propagation of
kink waves is most plausible source for observed Doppler shift
oscillations in spicules (Kukhianidze et al. 2005).

\begin{figure}
\includegraphics[width=9.5cm]{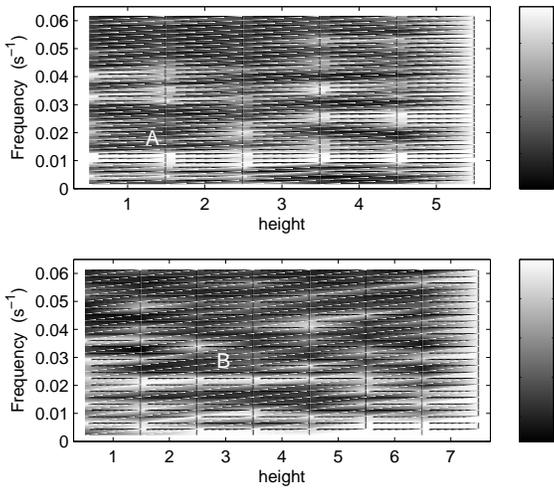}
\caption{Fourier power expressed in confidence levels as a function
of frequency and heights for the spicules II (upper panel) and IV
(lower panel). Light points correspond to higher power and dark
points correspond to lower one. The label 1 on the power scale
(right plots) corresponds to the 100$\%$ confidence level. On the
upper plot, we see long light feature located just above the
frequency 0.01 s$^{-1}$ (feature A). On the lower plot, there is
another long feature (feature B) located above the frequency 0.02
s$^{-1}$. \label{fig:epsart}}
\end{figure}

\begin{figure}
\includegraphics[width=9cm]{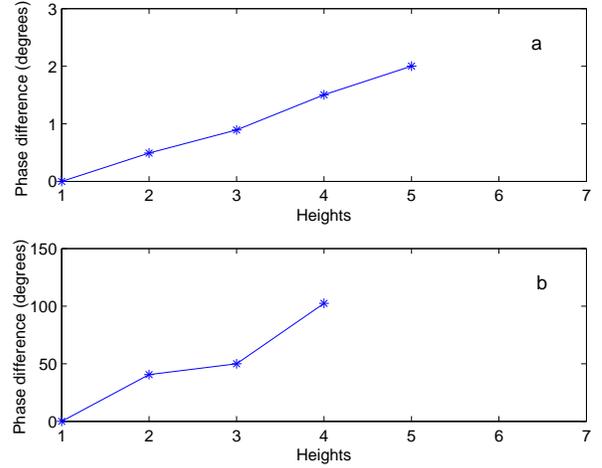}
\caption{Relative Fourier phase as a function of heights for
(a)$\sim$ 80 s period oscillation in spicule II (feature A on the
upper panel of Fig.7) and (b) 44 s period oscillation in spicule IV
(feature B on the lower panel of Fig.7). \label{fig:epsart}}
\end{figure}

Photospheric granulations may excite waves in anchored magnetic
tubes with two different time scales. The first time scale can be
similar to life-times of granular cells which gives wave periods of
$\sim$ 5-10 min. The second time scale can be related to spatial
scales of granular cells i.e. waves are excited with wave lengths
corresponding to cell spatial dimensions ($\sim$ 200-1000 km). Then
these waves may have periods of 20-100 s corresponding to typical
photospheric wave speed $\sim$ 10 km s$^{-1}$ (both sound and
Alfv\'en speeds probably have similar values in the photosphere).
Fig.6 shows that more than 2/3 of oscillation periods are located in
the range of 20-100 s, which clearly indicates to the granular
origin of the waves. It is seen that the periods are split into two
different ranges: 20-55 and 75-110 s. The reason of this phenomenon
is not clear. There are two potential candidates for this splitting:
either an unknown scaling of granular cells or fast and slow modes
in thin tubes. The first case implies the scaling of granular cells
into two different ranges of 200-500 and 800-1100 km, which is not
known to our knowledge (however probably it is interesting to look
into this problem in future). The second case implies the
propagation of fast and slow waves in spicules. There are two
different characteristic speeds of wave propagation in a magnetic
tube embedded in a field-free environment; kink  and tube speeds
(Edwin and Roberts 1983)
\begin{equation}c_k=c_A\sqrt{{{\rho_0}\over {\rho_0 +
\rho_e}}},\,\,\,c_T={{c_0c_A}\over {\sqrt{c^2_A + c^2_0}}},
\end{equation} where
\begin{equation}c_A={B_0\over {\sqrt
{4\pi\rho_0}}},\,\,\,c_0=\sqrt{{\gamma p_0}\over {\rho_0}}
\end{equation} are the Alfv{\'e}n and sound speeds inside
the tube respectively. Here $p_0$, $\rho_0$, $B_0$ are the pressure,
the density and the magnetic field inside the tube, $\rho_e$ is the
density outside the tube and $\gamma$ is the ratio of specific
heats. Fast waves propagate with a phase speed close to the kink
speed, while slow waves propagate with the tube speed. The density
is much higher inside the spicule than outside, therefore the kink
speed is close to the Alfv\'en speed. Let suggest that the Alfv\'en
speed is three times more than the sound speed in spicules at
observed heights (4000-8000 km) i.e. $c_A=3c_0$, then it implies
$c_k \approx 3c_T$. Then the fast waves propagate three times faster
than the slow waves. So if both waves are excited with the similar
wave length then the period of fast waves must be three times
shorter than that of slow waves. It is intriguing that the ranges
20-55 and 75-110 s have similar relation. Then the oscillations with
20-55 s period can be caused by the propagation of fast waves, while
the oscillations with 75-110 s can be due to the slow waves. It is
an interesting fact, that the oscillations with 75-110 s have
significant power, even more than ones with shorter period. The
oscillations may be signature of periodic vertical flows excited by
the resonant buffeting of granular cells (Roberts 1979). In this
interesting paper, Roberts suggested that the quasi-periodic
external buffeting of granules on magnetic tubes may lead to
periodic resonant vertical flows just as squeezing of a hosepipe
drives a jet of water. The period of the external forcing was taken
as about 5 min corresponding to mean life-time of granules.
Consequently, resonant periodic flows must propagate with the tube
speed and have the same $\sim$ 5 min period. However, there are
usually 3-4 granular cells in the neighborhood of each photospheric
vertical tube, which cause quasi-periodic squeezing of the tube.
Then, the mean period of external forcing will be 3-4 times shorter
than mean life time of granules. Consequently, the resonant periodic
flows will have 80-100 s period, which surprisingly coincides to the
observed periods. This phenomenon needs more vigorous analysis as it
may deal with spicule formation mechanism, but it is out of scope of
this paper.

Another interesting result is the clear peak at the period of 178 s
(Fig.6), which probably indicates to oscillations with a period of 3
min. Thus the 3-min oscillations may penetrate up to the heights of
4000-8000 km. It is interesting to check if 5-min oscillations are
presented in these heights. Unfortunately, statistical search of
this period in our data is restricted due to life times of spicules.

It is particularly important to understand whether the oscillations
are due to propagating or standing wave patterns. There are only 7-8
spatial points (corresponding to each heights) in our data,
therefore to infer the phase propagation is not easy. In the paper
I, we presented an illustrative example of phase propagation, but it
needs more vigorous treatment. The wave propagation can be revealed
through the variation of Fourier phase with position (Molowny-Horas
et al. 1997). Unfortunately, using the method in our data is
complicated as the oscillations have high confidence only at two
heights (third and fourth ones). The oscillations at other 6 heights
are not very confident due to poor fit of line profile to Gaussian.
However, some rough estimations still can be made. We calculated the
relative Fourier phase between heights for most pronounced features
(features A and B) of Fig.7. Fig.8 shows the relative Fourier phase
as a function of heights for (a) the feature A and (b) the feature
B. There is almost no phase difference between oscillations at
different heights for the feature A, which probably indicates a
standing-like wave pattern with period of $\sim$ 80-90 s. On the
contrary, there is the significant linear phase shift on the plot
(b), which indicates a propagating pattern with period of $\sim$
40-45 s. So the spicule II shows the standing-like pattern (or wave
propagation almost along line of sight, which seems unlikely) and
the spicule IV shows the propagation pattern. The propagation speed
for the feature B can be roughly estimated. The wave length can be
given as (Molowny-Horas et al. 1997)
\begin{equation}
\lambda = {{2\pi}\over {\Delta \phi}} \Delta s,
\end{equation}
where $\Delta \phi$ and $\Delta s$ are phase difference and distance
between first and last points. Then it gives the upper limit for
wave length as $\sim$ 6000 km. The actual wave length will be
shorter if spicule is tilted from the vertical. For the 35$^0$ tilt
(Trujillo Bueno et al. 2005) the actual wave length is 4800 km. Then
the phase speed of the feature B (with period of 44 s) can be
estimated as
\begin{equation}
\sim 110\,\, {\rm km \cdot s}^{-1},
\end{equation}
which is near the kink speed in the chromosphere. Thus, the feature
B is probably due to the kink wave propagation in the spicule IV.
The origin of the feature A is controversial. The oscillations are
almost in phase at 5 different heights, which indicates to a
standing-like pattern. The pattern can be formed due to the
reflection of slow tube waves at the transition to the
high-temperature corona. But the phase difference analysis is not
very confident in our case, therefore the results can be spurious.
Future detailed observations are necessary to confirm the
phenomenon.

\subsection{Spicule seismology}

Observed waves can be used to infer plasma parameters inside
spicules. The method called as {\it coronal seismology} is widely
used in coronal loops (Nakariakov and Ofman 2001) and prominences
(Ballester 2006).

In paper I we reported periodical distributions of H${\alpha}$
Doppler velocity with height. The spatial distribution has been
explained in terms of kink wave propagation from the photosphere
towards the corona. The wave length was estimated as ${\sim} 3500$
km. Then we may calculate wave phase speed with the help of observed
oscillation period, which enables to estimate a magnetic field
strength in spicules. Recently, Singh and Dwivedi \cite{sing}
estimated a magnetic field of 10-26 G in spicules. They used the
observed wave length from the paper I and the oscillation period of
1 min from Nikolsky and Platova \cite{nik}.

Fig.6 shows that most expected periods are $\sim$ 35 s and 90 s.
Then the phase speed can be estimated either as $\sim$ 100 km
s$^{-1}$ or $\sim$ 40 km s$^{-1}$. Both of them are higher than the
adiabatic sound speed for a temperature of $\sim$ 15 000 K at height
of $6000$ km (Beckers 1972), which gives $\sim$ 20 km s$^{-1}$.
Therefore we argue that observed spatial periodicity in Doppler
velocity with ${\sim} 3500$ km (paper I) is caused by the
propagation of kink waves with the period of 30-40 s. Then the kink
speed in spicules at the heights of 3800-8700 km from the
photosphere can be estimated as
\begin{equation}
c_k=90-115\,\, {\rm km \cdot s}^{-1}.
\end{equation}
Note, that the Fourier phase difference, roughly estimated for the
oscillation in spicule IV (feature B), also gives the similar value
of the phase speed ($\sim$ 110 km/s, see previous subsection).

As a plasma density is much higher inside spicules than outside,
then the kink speed is almost the Alfv{\'e}n speed. Expected
electron density in spicules at the height of 6000 km $8.9 \cdot
10^{10}$ cm$^{-3}$ (Beckers 1972) gives the plasma density of
$\rho_0 \approx 1.4 \cdot 10^{-13}$ g${\cdot}$cm$^{-3}$. Then, using
the kink speed (5), we may estimate the magnetic field strength in
spicules at the height of 6000 km as
\begin{equation}
B_0=12-15\,\, {\rm G}.
\end{equation}

This is in a good agreement with recently estimated magnetic field
strength in quiet-Sun spicules ($\sim 10$ G), which was obtained by
spectropolarimetric observations of solar chromosphere in the He I
$\lambda$10830 (Trujillo Bueno et al. 2005).

\subsection{Some critical points}

However, there are some critical points in interpretations of
observational data.

1. {\bf Fine structure of spicules}: it is quite possible that
spicules are consisted of several thinner magnetic tubes with
100-200 km diameter and this fine structure is not resolved due to
the spatial resolution of observations ($\sim$ 1 arc sec). Then each
spicule may consist of several {\it micro spicules}. Each micro
spicules may oscillate in its own way, which complicates the
spectral analysis of integrated Doppler shift.

2. {\bf Line of sight effect of several spicules}: some spicules
show a multi-component structure (i.e. several spicules are located
close together along a line of sight) mostly at lower heights, which
may give the impression of a velocity shift (Xia et al. 2005).
Therefore here only the spicules with well defined single-component
structure were chosen, however the line of sight problem can not be
completely excluded.

\section{Conclusion}

Here we analysed the old H${\alpha}$ spectra taken on the
coronagraph of Abastumani Astrophysical Observatory in solar limb
spicules. The time series include whole life times of four different
spicules at 8 different heights covering 3800-8700 km above the
photosphere (totally 32 time series). After DFT of the time series
we conclude:

\begin{itemize}

\item Doppler velocities of spicules undergo oscillations with periods of 20-110 s;

\item there are two different ranges of periods; almost half of the oscillatory periods are located in the range of
20-55 s and almost one third of oscillatory periods are located in
75-110 s;

\item most pronounced oscillation periods are $\sim 35$ and $\sim 90$ s;

\item there is an interesting peak at 178 s period, which is clear evidence of 3-min
oscillations at the heights of 3800-8700 km (detection of 5-min
period in our data is restricted due to life times of spicules);

\item we suggest that the oscillations are caused due to wave propagations in the chromosphere;

\item the waves or periodic flows likely are generated in the photosphere by granular
cells;

\item the two different ranges of periods can be explained either due to fast and slow waves or
due to some unknown spatial scaling of granular cells;

\item observations can be used to infer plasma parameters in spicules, i.e. {\it spicule
seismology};

\item estimated magnetic field strength in spicules is $12-15$ G
at the height of $\sim$ 6000 km above the photosphere.

\end{itemize}

However, some problems may arise in interpretations of observational
data related to spicule fine structure or multi-component nature,
which may lead to the impression of the velocity shift (Xia et al.
2005). Therefore future detailed spectroscopic observations with
higher spatial resolutions are needed to confirm high frequency wave
propagations in the chromosphere.

\section{Acknowledgements}

The work was supported by the grant of Georgian National Science
Foundation GNSF/ST06/4-098. A part of the work is supported by the
ISSI International Programme "Waves in the Solar Corona". We thank
Prof. Kiladze for helpful comments and the referee for stimulating
suggestions.

\end{document}